# The Emerging Planetary Nebula CRL 618 and its Unsettled Central Star(s)

B. Balick[1], A. Riera[2,3], A. Raga[4], K. B. Kwitter[5], and P. F. Velázquez[4]

[1] Department of Astronomy, University of Washington, Seattle, WA 98195-1580, USA; balick@uw.edu

[2] Departament de Física I Enginyeria Nuclear, EUETIB, Universitat Politècnica de Catalunya, Comte d'Urgell 187, 08036 Barcelona, Spain, angels.riera@upc.edu

[3] Departament d'Astronomia i Meteorologia, Universitat de Barcelona, E-08028 Barcelona, Spain

[4] Instituto de Ciencias Nucleares, Universidad Nacional Autónoma de México, 04510 D.F., México., raga@nucleares.unam.mx, pablo@nucleares.unam.mx

[5] Department of Astronomy, Williams College, Williamstown, MA 01267, USA; kkwitter@williams.edu

ABSTRACT

We report deep long-slit emission-line spectra, the line flux ratios, and Doppler profile shapes of various bright optical lines. The low-ionization lines (primarily [N I], [O I], [S II], and [N II]) originate in shocked knots, as reported by many previous observers. Dust-scattered lines of higher ionization are seen throughout the lobes but do not peak in the knots. Our analysis of these line profiles and the readily discernible stellar continuum shows that (1) the central star is an active symbiotic (whose spectrum resembles the central stars of highly bipolar and young PNe such as M2–9 and Hen2–437) whose compact companion shows a WC8-type spectrum, (2) extended nebular lines of [O III] and He I originate in the heavily obscured nuclear H II region, and (3) the Balmer lines observed throughout the lobes are dominated by reflected Hα emission from the symbiotic star. Comparing our line ratios with those observed historically shows that (1) the [O III]/Hβ and He I/Hβ ratios have been steadily rising by large amounts throughout the nebula, (2) the Hα/Hβ ratio is steadily decreasing while Hγ/Hβ remains nearly constant, and (3) the low-ionization line ratios formed in the shocked knots have been in decline in different ways at various locations. We show that the first two of these results might be expected if the symbiotic central star has been active and if its bright Hα line has faded significantly in the past 20 years.

Key words: planetary nebulae: individual (CRL 618) – stars: AGB and post-AGB – stars: winds, outflows

Section 1. Introduction

Proto-planetary nebulae ("pPNe") and their descendants, ionized planetary nebulae ("PNe") are formed from mass ejected in winds from an asymptotic giant branch (AGB) star. The high-resolution images (≤0."1) obtained over the past 20 years by the Hubble Space Telescope ("*HST*") have revealed that most pPNe show aspherical morphologies. About half of the imaged sample of 50 pPNe exhibit collimated bipolar or multipolar lobes (Sahai et al., 2011). As a result, collimated fast winds, bullets (aka clumps, knots), and thin cylindrical jets of modulated speeds have been proposed to operate during the brief post-AGB phase and be the primary agents for the shaping of PNe (Balick & Frank 2002, Sánchez-Contreras et al. 2002ApJ...578..269S, hereafter "SC+02", Balick et al. 2013ApJ...772...20B, hereafter "B+13", Riera et al. 2014A&A...561A.145R, hereafter "R+14", and references therein, submitted).

CRL 618 is one of the best sources to study stellar evolution from the AGB to the PN stages since it is bright, frequently observed, and well resolved by *HST* images (Trammell et al., 2002). It shows a broad array of emission lines (Goodrich 1991, Trammell et al., 1993ApJ...402..249T hereafter "TGD93") and large-scale proper motions (B+13, R+14) that allow us to trace its growth and evolution in some detail. Such studies suggest a kinematic age of about a century (e.g. SC+02, B+13, R+14). This and other evidence discussed below make CRL 618 a propitious target for probing the recent evolution of its central star as well as the origin of highly collimated late-life stellar outflows. We note for later purposes that TGD93 found that the permitted lines seen in the lobes are ≥15% polarized and, most likely, scattered into the slit by dust in the lobes. The source of this light in not obvious in HST images (B+13). It is likely to be the central nucleus (TGD93).

The rapidly evolving radio flux of the nuclear H II region and its other changing spectral characteristics have led several authors to suggest that the central star of CRL 618 is beginning a rapid traversal of the H-R diagram. Historically, CRL 618 has been brightening since its discovery (Walker & Price 1975). Gottlieb & Liller (1976) discovered that the B-band brightness of CRL 618 increased two magnitudes over 30 years (from 1940 to 1975). Kwok & Feldman (1981) first noted an increase in the radio continuum flux by a factor of two. The millimeter emission of the central HII region has also increased by a factor of 3.5 from 1977 to 1987 (Martín-Pintado et al. 1988). Such changes have been ascribed to an increasing flux of hydrogen-ionizing photons, $Q_H^o$, emitted by a post-AGB star evolving across the H–R diagram.

Recently Tafoya et al (2013A&A...556A..35T, hereafter "T+13") carefully compiled and analyzed the substantial, continuing, and highly systematic increases in the radio continuum flux and observed angular sizes of the core H II region since 1974. Among other properties, they estimated the onset of thermal radio emission from the obscured nuclear H II region to be 1971 ± 2. To quote from their paper, "we are witnessing the birth of a planetary nebula in CRL 618". T+13 argued that the observed time scales of the increasing radio flux mandate that the present mass of the central star must exceed ≈ 0.8 M$_\odot$ (Blöcker 1995). This requires that the initial mass of the star was initially ≈ 4 M$_\odot$, (late B-type), an unusually high value among PNe but marginally viable for the height of CRL 618 above the plane of the disk (≈120 pc).

The [O III] line has increased in flux since 1990. This line was not detectable in early observations reported by Goodrich (1991) and TGD93. In their extensive and methodical study of CRL 618 SC+02 noted an increase (decrease) of the [O III]/Hβ ([O I]/Hα) ratio by more than a factor of two over the course of a decade. Their preferred explanation for the changes in these line ratios was an increase in the shock speeds in the bright knots at the tips of the various outflow "fingers" (Fig. 1). However, using emission line ratios measured in 2003 by HST/STIS in the bow-shocks of CRL 618, Riera et al. 2011 analyzed newer HST/STIS long-slit spectroscopy to show that the shock velocities lie in a narrow range between 30 to 40 km s$^{-1}$, or more than a factor of two less than that estimated by SC+02. Their shock speed estimates agree with most earlier papers. If the [O III] line is not indigenous to the knots then another mechanism is needed to explain the evolution of the [O III]/Hβ ratio.

The primary purpose of this paper is to monitor the changes in optical lines and their ratios in order to test this hypothesis of increasing stellar ionization in detail. We obtained deep long-slit observations in two position angles passing across the nebular core and corresponding to the locations of three shocked fingertips (Fig. 1). The spectral dispersion was adequate to separate the knots from the emission elsewhere in the lobes of CRL 618 not only by location but also by the line broadening of bow shocks. This allows us to probe the changes of kinematics and to compare our results to those of SC+02 who observed with similar spatial and spectral resolution. While our data generally agree, our interpretation of them will turn out to be disparate in important ways.

The outline of this paper is as follows: In section 2 we present our data and describe the data acquisition and calibration. Section 3 is based on our spectral observations in order to identify the originating sources of the spectral lines and continuum, to interpret the complex Balmer decrement, and to describe the changes in Doppler shifts seen along our slits. In section 4 we combine our recently measured emission-line ratios with their historical counterparts in order to probe the evolution of various line ratios and their implications. In the final section we focus on the recent evolution of the optically obscured central star of CRL 618.

Section 2. Data, Calibration, and Reddening Corrections

The present observations were obtained with the APO 3.5-m telescope and Double Imaging Spectrograph (DIS) with a 1.″5 slit on the night of 13 February 2013. Slit placement is shown in Fig. 1. Blue (red) spectra cover 3996 to 5822Å (5669 to 6858Å) with an effective resolution of 1.5Å (70 km s$^{-1}$ at Hα and 90 km s$^{-1}$ at Hβ). Both slits included the nebular core, much of which is obscured by dust. The raw data were obtained in dark time under conditions of 1.″0-1.″5 seeing and photometric skies. The total exposure times were 96 min for the slit in P.A. = 94° (airmass 1.02 to 1.13) and 70 min for the slit in P.A. = 112° (airmass 1.19 to 1.59). Calibrations followed the tested procedures that we have traditionally used for DIS spectra (e.g., Henry et al. 2010) using G191B2B as a flux reference. The profiles of the bright lines are compiled in Fig. 2.

Spectra of the fingertips through the 1.″5 slit were then summed in various zones. Three zones of length = 3 pixels (2.″5) were used to produce a 1-D spectra at the locations of knot complexes E1-E2-E3, E4-E5, and W1-W2-W3 (see the grey boxes in Fig. 1 but bear in mind that seeing and atmospheric dispersion can scatter nearby light into the edge of the slit). The spectrum of the nucleus was extracted from two spectral rows containing the brightest lines from the feature identified as the "core" in Fig. 1. Gaussians were fitted to the extracted spectra within each box in order to derive line fluxes in each summation zone. The resulting line ratios are compiled in Table 1 along with corresponding results extracted from earlier publications in similar summation areas. All line ratios are uncorrected for reddening.

Section 3. Results from The Present Observations

In this section we interpret the spectral line ratios observed at APO in 2013.2. We will compare these results to historical line ratio measurements made with similar apertures by TGD93[1], Baessgen et al., 1997A&A...325..277B (hereafter "BHZ97"), Kelly et al., 1992ApJ...395..174K (hereafter "KLR92"), Baessgen et al., 1997A&A...325..277B (hereafter "BHZ97"), and SC+02 in the following section.

3.1 Three Classes of Emission

As Figs. 1 and 2 suggest, there are two distinct classes of emission lines, each of them distinguished by their positions along the slit and by the character of their line ratios. One of these is the obvious shock-excited knots where local low-ionization emission lines of [N I], [O I], [S II], and [N II] peak in surface brightness. These lines emitted in knots are relatively broad and Doppler shifted. The other major class consists of lines of "higher ionization" — notably Balmer, He I, and [O III] lines — that peak in the unresolved central core as well as features **A** and **B**. In addition we find a variety of faint [Fe III] emission lines that also peak in the core and otherwise follow the continuum light in Fig. 2. The [N II] lines have hybrid properties. They exhibit peaks in nearly all of these features.

Continuum light that is distributed amorphously and that contains very broad C III and C IV lines is also found centered at about 2″ east and west of the core. On the brighter eastern side of CRL 618 this light coincides with the relatively extended cyan region of Fig. 1 where light from the F547M (continuum) and F656N (H$\alpha$) filters both contribute.

3.2 The Central Stars

The very broad C III and C IV lines are far too wide (6–10Å) to be nebular (≤1Å). Using the WC classification criteria of Crowther et al. (1998), the ratios of the broad emission lines ratios are a good match to that of WC8-type nuclei (e.g., NGC40, Balick et al., 1996). Therefore original source of these lines and the continuum is likely to be an obscured WC star in the core. Apparently we observe its scattered light (along with nuclear continuum and emission lines) from patches of dust within the lobes.

The central core of CRL 618 is obviously an important region that deserves close attention. The *HST* image[2] (inset of Fig. 1) shows that the structure of the core is complex and that no star is readily visible. Light from the WC8 star does not appear to be visible in our spectra along this line of sight. However, a broadband F110W image (1.1 μm) in the *HST* archives shows a bright cool star and its diffraction spikes coincide (to within 0.″05) with the westernmost (green) peak of the core (inset, Fig. 1). The 1.1-

---

[1] The spectra of TDG93, KLR92, and BHZ97 were obtained at almost the same time and yet the values of H$\alpha$/H$\beta$, [N II]/H$\beta$, and [S II]/H$\beta$ derived by TGD93 ratios in the west lobe are larger than all of the others by a factor of three. These TGD93 ratios for the west lobe are ignored in this paper.

[2] Based on observations made with the NASA/ESA *Hubble Space Telescope*, and obtained from the Hubble Legacy Archive, which is a collaboration between the Space Telescope Science Institute (STScI/NASA), the Space Telescope European Coordinating Facility (ST-ECF/ESA), and the Canadian Astronomy Data Centre (CADC/NRC/CSA). The relevant GO programs are listed in Table 1 of B+13. Support for program GO11580 was provided by NASA through a grant from the Space Telescope Science Institute, which is operated by the Association of Universities for Research in Astronomy, Inc., under NASA contract NAS 5-26555.

μm peak presumably corresponds to a typical luminous ($\gtrsim 10^{3.5}$ $L_\odot$) and cool (2500 < $T_{eff}$ < 4000 K) post-AGB star whose character can be discerned from its spectrum.

Our spectrum of core shows the set of high-ionization lines noted earlier plus a series of [Fe III] lines, all of which peak within 1″ of the same location. This line set is an extremely close match to the spectra of the much brighter central stars of M2–9 and M1–91 (aka Hen2-437) observed at higher dispersion by Torres-Peimbert, Arrieta, & Bautista (2010RMxAA..46..221T). (See also section 3.4 below.)

For all of these reasons we posit that the central star of CRL 618 consists of a post-AGB—WC8 (±1) binary. The cooler star belongs to the class of stars at the cores of highly bipolar symbiotic PNe as M2–9, M1–91, BI Cru, Hen 2–104. Widened, luminous Hα lines dominate the emission-line spectra of most of these stars, so it seems plausible that much of the Hα in CRL 618 elsewhere along our slits (outside of the shock-excited knots) is stellar in origin. This is supported by the polarization of the Hα line found by TGD93.

B+13 found conspicuous and extended illumination cones of scattered continuum in their F606W, F110W, and F160W *HST* images. Moreover, *HST* images in the Hα line share the same light distribution (outside of the knots). Although the lines are faint, our spectra show extended [Fe III] emission in the lobes whose brightness distribution is very similar to continuum light. For these reasons we surmise that the Hα line, the stellar [Fe III] lines and the C III, C IV lines, and the stellar continuum light originate from the central stars and scatter from dust in the lobe into the line of sight.

3.3 The High-Ionization Nebular Lines

A comparison of the profiles of the [O III] and He I lines in Fig. 2 show that outside of the nucleus itself they too follow the distribution of continuum light. Therefore, leaving aside the locally shock-heated knots, we expect that much or most of the [O III] and He I lines observed in our slits in the lobes originate in the nuclear region — most likely from the optically invisible core H II region (T+13) and not within the lobes. Also just like the stellar Balmer lines and the continuum, the profiles of the [O III] and He I lines all show that their light freely escapes all the way to angular radii of ±10″ (see also SC+02).

We note from Fig. 2 that little of the [O III] flux arises in any of the shocked knots on the eastern side of the core. Consequently we agree with R+11 that the corresponding shock speeds are <50 km s$^{-1}$ at those positions. The shock speeds in the western fingertip may be larger since the [O III] lines show a wide redshifted peak at the position of knot W1.

3.4 Reddening

Nebular reddening is generally derived from the Balmer decrement. SC+02 argued that reddening measured from the Hα/Hβ line ratio has a very strong dependence on position. However, TGD93 noted that the reddening cannot be well determined from Balmer-line ratios unless the scattered light is removed first.

Based on our values of Hα/Hβ of 2013.2 we derive nominal extinction coefficients c(Hβ) = [log(Hα/Hβ) – log(Hα/Hβ)$_{c(H\beta)=0}$]/0.35 of 0.9, 0.8, 1.0, and 1.6 for the knot complexes E1-E2-E3, E4-E5, W1-W2-W3, and the core, respectively. SC+02 (Fig. 10) showed from their spectra that c(Hβ) peaks in the center of CRL 618 and drops with angular radius to the shock-excited knots – a trend that we confirm.  However, our estimate of the central Hα/Hβ ratio (c. 2012.3) is 60% lower than theirs (c. 2001.8).  In addition we find that the Hα/Hβ ratio seen in the fingertips (Table 1) also declined by 60% since their measurements.  This evolution of c(Hβ) measured in this manner is not expected on time scales of ten years.

There are two plausible explanations for the odd behavior of c(Hβ) with position and time.  Suppose, as we asserted earlier, Balmer lines emitted from the nebula and/or the star deep inside the obscured nuclear zone of CRL 618 are scattered into our line of sight as well as absorbed by dust in the lobes.  Suppose also that the scattering and absorbing dust is most concentrated near the nebular center (as is clearly the case).  Then we can expect the Balmer line ratios of the detected light to change rapidly with distance from the nucleus.  This is certainly the case in M2–9 (M1-91) where Torres-Peimbert et al. (2010) found Hα/Hβ = 17.2 (7.5) and Hγ/Hβ = 0.53 (0.46) near their nuclei (and where the emerging stellar Hα lines are 15–20Å in width) and where the line reddening drops further into the lobes.  If the central star of CRL 618 shares these intrinsic Balmer line ratios then nebular reddening is indeterminate from regions heavily influenced by the stellar spectrum.  Note also that the secular decrease observed in the Hα/Hβ ratio (section 4) seen in our summation zones would be attributable to changes of the spectrum of the symbiotic star.  Such changes are not unusual.

Alternately, if the lines originate in a dense nuclear H II region whose density $> 10^5$ cm$^{-3}$ then the Balmer decrement — especially the Hα/Hβ ratio — will be enhanced by collisional excitation from the ground state (Netzer 1975).  Densities of this magnitude are certainly found within ±0.″5 of the nucleus (T+13) and the outer knots (R+11).  Thus we expect relatively large ratios of Hα/Hβ near the nucleus and again in the fingertips.  On the other hand any Balmer lines emitted within the lobe interiors (where the densities are well below $10^5$ cm$^{-3}$) will have more normal nebular ratios (Hα/Hβ ≈ 2.8).  In this case Hα/Hβ and c(Hβ) will peak at the nucleus and again at the fingertips.  This prediction is not consistent with our long-slit observations or those of SC+02 and seems untenable.

Given this complex discussion of origin of the Balmer lines in our spectra we decided not to correct our line ratios in Table 1 for reddening.  That turns out to be unimportant since our goal is to search for *changes* of line ratios over time.  This can be achieved directly from calibrated line ratios with no reddening corrections.  What's more, it is the only sensible strategy given the huge disparities in the values of c(Hβ) that were adopted by the authors of earlier references.

This is not to deny the existence of reddening, only the methodology of its derivation. To judge from Table 1 and Fig. 4, the Hγ/Hβ ratio seems to be uniform in both fingertips and invariant in time.  On the other hand, the stellar Hγ and Hβ lines are relatively faint, so the use of their ratio is not a reliable means for estimating the nebular reddening.  Moreover, errors in c(Hβ) could have dramatic effects on the ratios of red nebular lines to Hβ. (Note that we do not include Hδ in this analysis because its flux is weak.)

3.5 The Low-Ionization Nebular Lines

The present measurements are compiled in Table 1. Other than [N II], these lines do not arise in the nuclear spectrum, so they are seemingly intrinsic to the shocked knots. The line ratios at various past epochs have been fully analyzed in other papers (see R+11).

3.6 Features *A* and *B* in Fig. 2

Both **A** and **B** are conspicuous features in our observations (Fig. 2) that were previously identified as a pair of features labeled "B" by SC+02 (see their Fig. 4). **A** and **B** share these properties: like the scattered light from the core, both **A** and **B** are visible only in lines of higher ionization. They lie on opposite sides and within ≈2″ of the nuclear H II region.

The similarities end there. **A** has the same Doppler shift as the eastern lobe as a whole. **A** is visible in the [N II]5755Å panel of Fig. 2 but not in the [N II]6583Å panel. It appears to coincide with the brightest continuum seen in the slit and with amorphous cyan feature in Fig. 1. Thus **A** is likely to be a dense dust clump within the eastern lobe that scatters light from the central star and the dense, compact nuclear H II region (T+13).

The nature of **B** is more enigmatic. **B** lies on the western edge of the dark dust lane. Like the knots to its west, its emission lines are broad, redshifted and comparable in flux. However, unlike the western knots, **B** is visible only in the Balmer, He I, and [O III] lines. We do not see any obvious spatial counterpart to **B** in Fig. 1 (c. 2009.6).

3.7 Kinematic Trends

We noted in section 1 that the proper motions of sharp-edged knots in CRL 618 show systematically increasing displacements of the knots with offsets from the nucleus. As a result B+13 and R+14 argued that the knots share a common kinematic age of about a century.

However, whether this same ballistic expansion pattern applies to the lobe interiors is much less clear, as an inspection of the slit profiles in Fig. 2 show. A single ejection age for the knots and the lobes requires that all of their profiles show the same measurable slopes of Doppler shift and distance. Yet the global pattern of motions of lobes and knots are distinct. Fig. 2 clearly shows that the Doppler trends of the knots with offset are steeper than the Doppler shifts of the more amorphous scattered high-ionization lines throughout the lobes and extending well past the fingertips. Thus the kinematic age of the scattering dust is longer than that of the knots. B+13 also argued that the rings of reflecting dust surrounding the bright fingers of CRL 618 expand as one unit and the knots as another. They asserted that the kinematic ages of the rings are ≳4–5 times greater than those of the knots.

3.8 Other Results from our Spectra

In principle the densities in the knots can be determined from the ratios of the [S II] 6717 and 6731Å lines. However, like many other previous observers, we find that the ratio of the density-sensitive [S II]6717Å and 6731Å lines is 0.52 ± 0.02; that is, very close to the density-saturation limit of $n_{[S\,II]} \approx 20{,}000$ cm$^{-3}$. Other evidence discussed by SC+02 and R+11 shows that $n_{[S\,II]}$ *probably* exceeds $10^5$ cm$^{-3}$. The unusually large [N II]5755Å/(6548+6583Å) ratio seen in our nuclear spectrum supports their conclusions.

We are able to measure the temperature-diagnostic ratio [O III]4363/(4959+5007)Å from our spectra but with 30% uncertainty (assuming $c(H\beta) = 0$) and no collisional de-excitation of the nebular [O III] lines. The ratio in the fingertips averages 0.054, a result consistent with an [O III] temperature, $T_{[O\ III]}$, of $1.7 \times 10^4$ K. However, $T_{[O\ III]} > 20{,}000$ K if $c(H\beta) = 2$, as reported in some other papers. If the [S II] density also applies to the O++ zone in the nuclear H II region, as found by T+13, then the [O III] nebular lines can be partially quenched by collisions. This renders our value of $T_{[O\ III]}$ an underestimate. All considered, the temperature estimated from radio continuum observations, 13,000 K (Martín-Pintado et al., 1988), is more plausible than ours.

Section 4. Results from Historical Line Ratios

CRL 618 is a nebula with a complex morphology that changes quickly and a nuclear H II region whose radio size and flux are steadily increasing. In this section we shall explore the evolving character of its optical spectrum observed in zones proximate to the spectral summation regions that we used for the analysis in section 3. The relevant data are compiled in Table 1 and plotted in Fig. 4.

By way of background, T+13 have analyzed radio flux-density data spanning 40 years. During that time the 5-GHz thermal radio flux has increased by more than a factor of 5 and its angular size has doubled to 0."5. T+13 asserted that that the stellar flux of ionizing photons, $Q_H^o$, has multiplied, the stellar mass loss rate is presently $10^{-5.2}$ $M_\odot$ $y^{-1}$, the shape of the thermal radio spectrum (that is, the nebular emission measure) is not changing, and that the nebular density drops from $10^7$ cm$^{-3}$ in the core to $10^5$ cm$^{-3}$ at the nebular edge (implying a very steep radial thermal pressure gradient within the H II region).

4.1  Changes in [O III]/Hβ and He I/Hβ Lines from the Core

Changes in the ratios of higher-ionization lines, [O III]5007Å/Hβ and He I5876Å/Hβ are largely ascribable to changes in the optical spectrum emitted by the star and the compact H II region around it. Inspection of Table 1 and Fig. 4 (blue lines) shows that the [O III]5007Å/Hβ ratio has increased dramatically on opposite sides CRL 618 from almost zero in the early 1990s (Goodrich 1991, TGD93 Figs. 1 & 3) to ≈30% today (±5%) in knots E1–E3 and W1 and W2, over 50% in knots E4+E5, and 75% in the nuclear region. Although the data are lower in quality, the same secular rise applies to the HeI5876Å line.

The rises in the [O III]/Hβ and He I/Hβ ratios are likely to be direct outcomes of changes in the spectrum that is emerging from the nuclear H II region. As noted earlier, scattered nuclear He I and [O III] also nicely explains the narrow line widths and constant Doppler shifts of the observed line profiles of these lines that extends beyond the fingertips of CRL 618 (Fig. 2) where the dust is relatively stationary.

The secular changes in the ratios suggest that the effective temperature, $T_{eff}$, of stellar photons above 1 Ryd (equivalently, the ratios $Q_O^+/Q_H^o$ and $Q_{He}^o/Q_H^o$) have increased steadily over the previous three decades. Models of UV radiation transfer suggest that the effective stellar temperature, $T_{eff}$, now exceeds ≈ 30,000-35,000 K if all of the stellar ionizing photons are absorbed within the nebula (Dopita and Sutherland 2003, Fig. 9.4).

This temperature is roughly consistent with late-type WC central stars whose nebulae show [O III]5007Å/Hβ ≲ 1 (Balick et al. 1996).

Alternate explanations are also viable. Another mechanism that will increase the emissivity of the $^1D_2$ [O III] lines relative to Hβ is a decrease in local density below $10^6$ cm$^{-3}$ that favors radiative over collisional de-excitation rates in O$^{++}$. However, this won't account for changes in the He I/Hβ ratio. Also, although we find no direct evidence to favor the hypothesis, we cannot exclude a decline in the Hβ flux as a contributing cause.

4.2 Gross changes in Balmer line ratios since 1990

In normal ionized nebulae the ratio of Balmer lines is sensitive almost exclusively to foreground reddening. Consequently the Balmer line ratios are not expected to change on time scales of a decade. As we shall show, CRL 618 is a standout exception.

The historical Balmer line ratios are compiled in Table 1 from spectrophotometric measurements since 1990. For consistency we consider only the total (polarized plus unpolarized) line ratios obtained by TDG93 through their 2″ slit. (However, as noted earlier, we shall ignore TGD93's red-line ratios in the western lobe since they differ incongruously from other contemporary measurements.)

A quick glance at Table 1 and Fig. 4 (black lines) tells a story of significant temporal changes in the Hα/Hβ ratio. Hα/Hβ decreased steadily by factors of 2 between 1994 and 2013. During this time the Hγ/Hβ ratio did not change appreciably. Since the changes in the Hα/Hβ ratio pertains to both lobes there is likely to be a common cause, namely changes in the emergent fluxes of its Balmer lines (see also section 3.4).

A stellar origin for the scattered Hα line of CRL 618 is confirmed by its modestly broader width (≈200 km s$^{-1}$) relative to other lines in its spectrum (≈60 km s$^{-1}$; see SC+02 Fig. 8). Many symbiotic stars are characterized by bright and broad Hα lines (e.g., Munari & Zwitter 2002A&A...383..188M). In addition, erratic changes in the fluxes and profile shapes of Balmer lines are not unusual for some well-studied symbiotic systems such as *CH Cyg* where good data are available (Burmeister & Leedjärv 2009A&A...504..171B, esp. Figs. 3 & 5).

Accordingly we propose that the variations of the Hα/Hβ ratio outside of the knots in CRL 618 can be traced back to changes in the Balmer line fluxes of the stellar spectrum. This hypothesis was investigated using *HST* images in the F656N filter spanning 10.8 years to search for differences in the absolute Hα line surface brightness on the relatively bright eastern side of CRL 618 (Fig. 3)[3]. The dark-grey residuals in the difference image are indicative of a general decrease in the dust-scattered Hα flux. All of the bright stationary features near the core faded by a factor of about 0.8 ± 0.2 except for the stellar peak on the west edge of the core whose flux increased by a factor of 1.9 ± 0.2. The overall fading of Hα by about 20% appears congruous with the secular decrease of the Hα/Hβ ratio found in the spectroscopic measurements of Table 1.

---

[3] The archival images have units of detected count rates. The displayed images were adjusted for the factor of 2.36 increase in relative system efficiency of WFC3 relative to WFPC2.

A nuclear origin applies also to the narrow lines observed at the nuclear Doppler shift in the spatial extensions of the other higher-ionization lines beyond the fingertips.

4.3 Changing Ratios of Low-Ionization Lines

Table 1 and Fig. 4 present changes in the ratios emitted by neutral and singly-ionized metals, [NI]$\lambda\lambda$5198+5200Å/H$\beta$, [OI]$\lambda\lambda$6300+6363Å/H$\beta$, [S II]$\lambda\lambda$6717+6731Å/H$\beta$, and [N II]$\lambda\lambda$6548+6583Å/H$\beta$. The ratios decreased far more substantially in the western knots than in the eastern. That is, there is no reason to attribute the changes to a common or nuclear mechanism especially in light of the large knot-to-knot temporal brightness and shape variations found by R+14 and the highly random brightness changes of the fingertips seen by B+13 in multi-epoch *HST* images. We cannot exclude (unlikely) increases in the H$\beta$ flux or changes in reddening as contributing causes of ratio changes.

Section 5: Conclusions

The primary new scientific findings of this paper are these:

• Although no central star is directly visible at optical wavelengths, we detect its dust-scattered light in the lobes of CRL 618. In addition to continuum, the scattered light contains very conspicuous non-nebular features of a WC8-type star plus the continuum and Balmer and [Fe III] lines of a cooler star. The nuclear optical spectrum matches the line set found in bright symbiotic stars of several PNe with highly bipolar morphologies. Also, a very cool luminous stellar object seen in *HST* images at 1.1 and 1.6 µm coincides with the western edge of the nuclear zone and the eastern edge of a deep dust lane. By analogy with other symbiotics, the apparent changes in scattered H$\alpha$ flux are best explained if the H$\alpha$ flux predominantly arises in the symbiotic star.

• Using our data in combination with HST images indicates that the point-like and extended [O III] and He I lines are emitted in the same nuclear H II region that has brightened and expanded in the past 40 years (T+13). The [O III]/H$\beta$ and He I/H$\beta$ line ratios have increased by about the same factor as the thermal radio continuum flux from the nuclear H II region. Multi-epoch *HST* images in the H$\alpha$ line indicate that emission lines from the nuclear H II region may be visible optically along the line of sight near the location of the cool star.

• The nebular Balmer decrement has long been known to be peculiar. If the central star is a symbiotic, then much of the flux of H$\alpha$ and possibly some of the other Balmer lines that we observed is likely to be scattered starlight. This is amply confirmed in the line polarization studies of TGD93. Symbiotics often have complex Balmer profiles shapes that evolve over time—as do the forbidden lines seen in their spectra. Thus the observed secular changes in the [O III]/H$\beta$, He I/H$\beta$, and H$\alpha$/H$\beta$ ratios may simply reflect the erratic and poorly understood behaviors of an active symbiotic star.

• The emission lines of relatively low ionization arise mostly in shock-excited knots with Doppler broadened bow shocks that have been studied in previous papers. Locally produced [O III] lines do not arise in the knots except (perhaps) in knot W1. Thus the shock speeds in the other knots are likely to be $\lesssim$50 km s$^{-1}$, far slower than the observed radial proper motions of the knots in the past 20 years. It is possible that some of the line emission arises in the reverse shock propagating into a clump or jet.

• The steady increases in the thermal radio flux and the nebular [O III]/Hβ and He I/Hβ ratios may imply that the ionizing star is becoming more luminous and hotter. Possibly its effective temperature has been increasing in the past 40 years, though the observed evolution time scale seems unreasonably short unless the core mass of the WC8's progenitor star is exceptionally large ($\gtrsim 4\ M_\odot$). As noted by Burmeister & Leedjärv (2009) in their study of the spectral changes of the symbiotic *CH Cyg*, fluctuations in the UV flux may also be connected with thermal instabilities in the accretion disk of the ionizing companion star as its mass accumulates from its bloated companion star.

T+13 found that densities in the nuclear H II region decreases sharply with position to $10^5$ cm$^{-3}$ at its outer edge. The recombination and cooling times of gas at this density are a few months or less. Therefore the thermal radio flux and the ionization structure of the nuclear H II region will track any decrease of stellar $Q_H^o$ in the future. Such changes are expected for symbiotics but not for normal stars as they traverse the post-AGB tracks in the H–R diagram. Thus continued monitoring of the radio flux of the nuclear H II region of CRL 618 can help to clarify the nature of its central ionization source.

<span style="color:green">Large secular changes in the [O III]/Hβ ratio have been seen around the dusty Mira–white-dwarf symbiotic star V1016 Cyg (1965 eruption), the dusty symbiotic nova HM Sge (1975 outburst), as well as few compact and rapidly evolving bipolar PNe such as Hb 12 and IC 4997 (Gutiérrez-Moreno, Moreno, & Cortés, 1995). These authors show that the [O III]/Hβ ratio of HM Sge increased fourfold between 1976 and 1988.</span>

Much or all of the optically visible nebula of CRL 618 is the outcome of a stellar outburst a century ago. Despite many careful observations of the nebula the active mass ejection process remains unclear. B+13 argued on very general physical grounds that whatever the original ejection mechanism may have been, the formation of high-speed clumps was far more likely than jets or winds for explaining the shapes of CRL 618's thin multiple fingers since highly collimated flows tend to become unstable at the collimation point.

More recently R+14 and Velázquez et al. (2014, submitted) have successfully modeled the evolving morphology of CRL 618 by assuming a jet with periodic (time-dependent) ejection speed that is launched from the accretion disk around a compact companion that orbits a nearby mass donor in an eccentric orbit. The correspondence of the stars in this model to WC8 and cool giant stars of CRL 618 is obvious, and the bright nuclear [Fe III], He I, and [O III] lines suggest that mass transfer is in an active state. The ejection model requires that the kinematic ages of shocked knots at the heads of the intermittent jets is a monotonic function of their distance from the nucleus. This is consistent with the wavelet analysis of the proper motion studies by R+14.

Whatever else we may have presented in this paper, it is clear that a binary system lies at the heart of CRL 618 and its complex and rapidly evolving morphology, spatial growth, and spectral evolution. The Doppler shifts of its narrow [Fe III] lines should be monitored to see whether the period and eccentricity the binary orbit can be established. Complementary x-ray observations can clarify the role of fluctuations or instabilities in the mass lying within the inner accretion disk of the compact companion star.

ACKNOWLEDGEMENTS

We are pleased to thank the staff of Apache Point Observatory for their help with the observations. We thank Daniel Tafoya for helpful comments. B.B. is grateful to the NSF for support under grant AST- 0808201. This research has made extensive use of NASA's Astrophysics Data System. ACR and pFV acknowledge financial support from CONACyT grants 167611 and 167625 as well as DGAPA PAPIIT grant IG100214 (Mexico). A.R. is supported by grant MINECO AYA2011-30228-CO3 (Spain).

This project was supported in part by the National Science Foundation from grant AST-0808201 and by *HST* GO grant 11580. Support for GO11580 was provided by NASA through a grant from the Space Telescope Science Institute, which is operated by the Association of Universities for Research in Astronomy, Incorporated, under NASA contract NAS5-26555.

Some of the data presented in this paper were obtained from the Multimission Archive (MAST) at the Space Telescope Science Institute (STScI). STScI is operated by the Association of Universities for Research in Astronomy, Inc., under NASA contract NAS5-26555. Support for MAST for non-HST data is provided by the NASA Office of Space Science via grant NAG5-7584 and by other grants and contracts.

Facility: APO (DIS), HST (WFC3, WFPC2)

REFERENCES

Baessgen, M., Hopfensitz, W., & Zweigle, J. 1997A&A...325..277B (BHZ97)
Balick, B. & Frank, A. 2002ARA&A..40..439B
Balick, B. Huarte-Espinosa, M, Frank, A. Gomez, T., et al. 2013ApJ...772...20B (B+13)
Balick, B., Rodgers, B., Hajian, A., Terzian, Y., & Bianchi, L. 1996AJ....111..834B
Blöcker, T. 1995, A&A, 299, 755
Burmeister, M. & Leedjärv, L. 2009A&A...504..171B
Crowther, P.A., De Marco, O., & Barlow, M.J. 1998MNRAS.296..367C
Dopita, M.A. & Sutherland, R.S., Astrophysics of the diffuse universe, Berlin, New York: Springer, 2003. Astronomy and astrophysics library (ISBN 3540433627), page 239.
Goodrich, R.W. 1991, ApJ...376..654G
Gottlieb, E.W. & Liller, W. 1976ApJ...207L.135G
Gutiérrez-Moreno, A., Moreno, H., & Cortés, G. 1995PASP..107..462G
Henry, R.B.C., Kwitter, K.B., Jaskot, A.E., Balick, B., Morrison, M.A., & Milingo, J.B.2010ApJ...724..748H
Kelly, D.M., Latter, W.B., & Rieke, G.H. 1992ApJ...395..174K (KLR92)
Kwok, S. & Feldman, P.A. 1981ApJ...247L..67K
Martín-Pintado, J., Bujarrabal, V., Bachiller, R., Gomez-Gonzolez, J., & Planesas, P. 1988A&A...197L..15M
Munari, U. & Zwitter, T. 2002A&A...383..188M
Netzer, H. 1975MNRAS.171..395N
Riera, A., Raga, A.C., Velázquez, P.F., Haro-Corzo, S.A.R., & Kajdic, P. 2011A&A...533A.118R (R+11)
Riera, A., Velázquez, P.F., Raga, A.C., Estalella, R., & Castrillón, A. 2014A&A...561A.145R (R+14)
Sánchez Contreras, C., Sahai, R., & Gil de Paz, A. 2002ApJ...578..269S (SC+02)
Sahai, R., Morris, M.R., & Villar, G.G. 2011AJ....141..134S
Tafoya, D., Loinard, L., Fonfría, J.P., & Vlemmings, W.H.T. 2013A&A...556A..35T (T+13)

Torres-Peimbert, S., Arrieta, A., & Bautista M. 2010RMxAA..46..221T
Trammell, S. R., & Goodrich, R. W. 2002ApJ...579..688T (TG02)
Trammell, S.R., Dinerstein, H.L., & Goodrich, R.W. 1993ApJ...402..249T (TGD93)
Walker, R.G. & Price, S.D. 1975, AFCRL Technical Report 75-0373

Central Stars of CRL 618 (Balick et al.), draft 8 July 2014    page 14

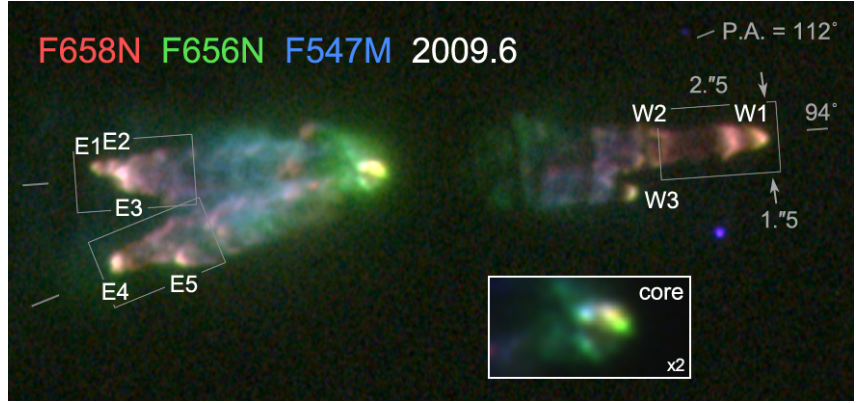

Fig. 1. An overlay of F658N, F565N, and F547M HST images from 2009.6. These filters are dominated by [N II], Hα, and a combination of [NI] and stellar continuum, respectively. Knots of particular interest in this paper and the approximate regions within which our APO spectra of 2013.2 were integrated are shown. The inset shows a magnified rendition of the core region in the same filters. All images were normalized to their peak values before use in this display. Color intensity is proportional to the log of surface brightness.

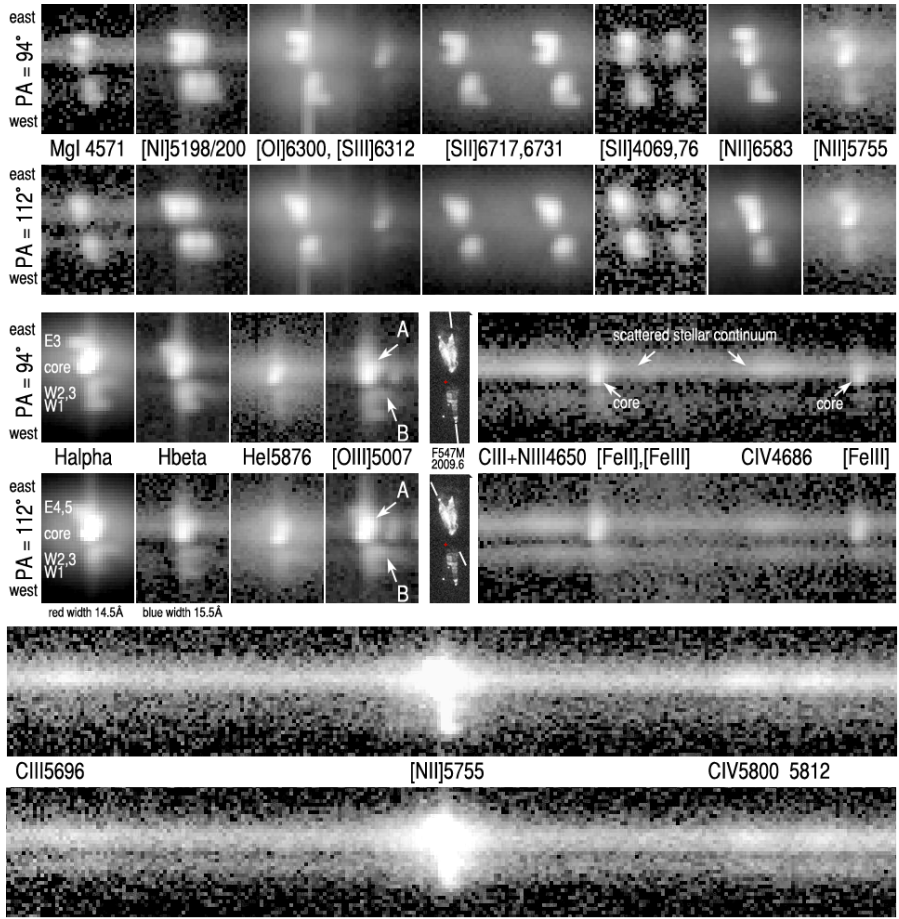

Fig. 2. Logarithmic intensities in spectral regions around brighter emission lines. The lines are shown in pairs with spectra in the slit at P.A. =94° at the top and P.A. = 112° below them. Line identifications are shown between the pairs. The F547M HST image (continuum plus [N II]5198+5200Å) is shown at the center of the figure with white lines showing the loci of the center of the slits. Most frames are 100 pixels in width (15.5Å for blue DIS spectra and 14.5Å for red DIS spectra) and 140 pixels in height. Sky lines of [N I] and [O I] are visible beyond the limbs of the fingers in this high-contrast display.

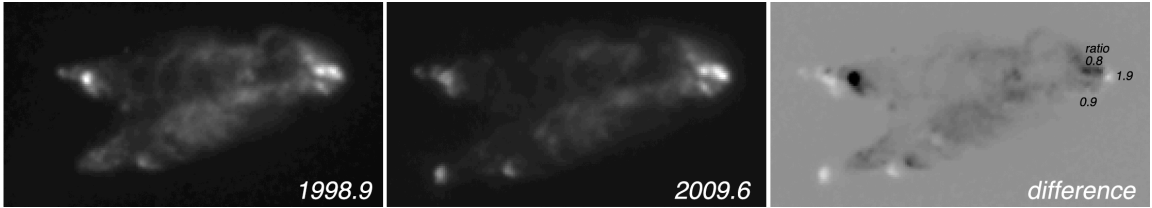

Fig. 3. Linear displays of HST F656N (Hα) images of eastern half of CRL 618 taken at the indicated epochs, regridded to a common pixel size, and displayed in units of detected counts per pixel per second. The left (middle) image is a WFPC2/PC (WFC3) image with each camera's F656N filter. The grey levels have been adjusted for a factor of 2.36 difference in camera total throughputs. The right panel is the difference image after alignment. The ratio of two-epoch count rates for selected core features are shown in small black numerals.

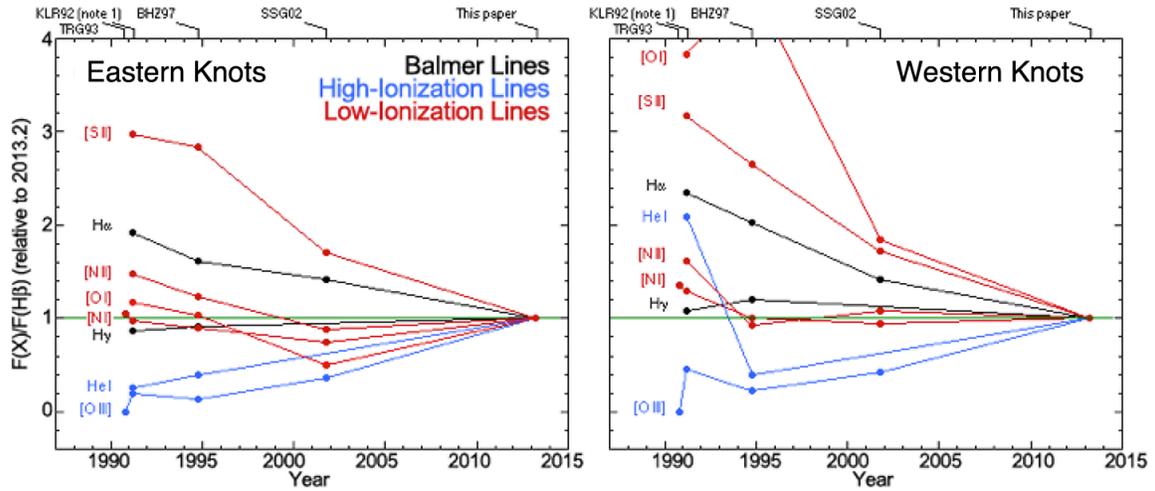

Fig. 4. Changes in line flux ratios relative to the data obtained in PA=94° during 2013.2. The colors indicate certain groups of lines discussed in the text, as indicated in the left panel. Some of the scatter is measurement uncertainty but the flux summation areas and slit geometries areas used by various authors also play a role. Note 1: The flux summation area is not described in the KLR92 paper.

Table 1. Measured Emission-Line Ratios (Hβ = 100) at Various Slit Locations*

| Reference | TDG93 | KLR92 | BHZ97 | SC+02 | This paper | TDG93 | KLR92 | BHZ97 | SC+02 | This paper | This paper | This paper |
|---|---|---|---|---|---|---|---|---|---|---|---|---|
| Slit width | 2."0 | 2."5 | 1."2 | 2."5 | 1."5 | 2."0 | 2."5 | 1."2 | 2."5 | 1."5 | 1."5 | 1."5 |
| P.A. | 90 | 90 | 90 | 93 | 94 | 90 | 90 | 90 | 93 | 94 | 112 | 94 |
| Epoch | 1990.8 | 1991.2 | 1994.8 | 2001.8 | 2013.2 | 1990.8 | 1991.2 | 1994.8 | 2001.8 | 2013.2 | 2013.2 | 2013.2 |
| Slit Location | East Lobe (E1-E2-E3 complex) | | | | | West Lobe (W1-W2-W3 complex) | | | | | E4 + E5 | Core |
| [SII] 4068.6 | | 42 | 30.9 | | 54.8 | | 77 | 56 | | 37.6 | 6.0 | |
| [SII] 4076.3 | | 13.5 | 10.9 | | 18.2 | | 10.8 | 17 | | 12.7 | 2.1 | |
| Hδ 4101.7 | | 2.0 | 17.3 | | 19.4 | | 0 | 15.4 | | 9.9 | 8.6 | 8.2 |
| Hγ 4340.5 | | 31.3 | 33.1 | | 35.9 | | 29.4 | 33 | | 27.3 | 21.6 | 29.6 |
| [OIII] 4363.2 | | | | | 2.9 | | | | | 2.1 | 3.2 | 4: |
| MgI] 4571.1 | | 11.1 | 8.9 | 7.6 | 15.9 | | 17.7 | 19 | 10.9 | 12.1 | 4.6 | (a) |
| Hβ 4861.3 | 100 | 100 | 100 | 100 | 100 | 100 | 100 | 100 | 100 | 100 | 100 | 100 |
| [OIII] 4958.9 | unseen | 2.1 | | 4.3 | 10.8 | unseen | 3.9 | | 3.6 | 8.4 | 17.5 | 22.0 |
| [OIII] 5006.8 | unseen | 6.7 | 4.9 | 13 | 35.1 | unseen | 12.7 | 6.2 | 11.8 | 28.1 | 56.4 | 75.1 |
| [NI] 5198+5200 | 105 | 97 | 89 | 74 | 99.5 | 199 | 191 | 148 | 139 | 147 | 103 | (a) |
| [NII] 5754.6 | 23 | 21.2 | 22.9 | | 26.4 | 20 | 21.6 | 27 | | 2.1 | 12.6 | 20 |
| HeI 5875.6 | | 9.0 | 13.3 | | 34.2 | | 8.8 | | | 4.2 | 18.9 | 47 |
| [OI] 6300.3 | 548 | 478 | 420 | 204 | 406 | 1192 | 1039 | 1240 | 500 | 271 | 201 | (1) |
| [SIII] 6312.1 | | 9.1 | | 5.4 | 4.8 | | | 2.5 | | | 9.5 | 20: |
| [OI] 6363.8 | 189 | 161 | 141 | 72 | 124 | 438 | 379 | 424 | 164 | 92.2 | 70.6 | (a) |
| [NII] 6548.1 | 65 | 69.6 | 53 | 35 | 45 | 502(b) | 150 | 83 | 91 | 84.3 | 34.7 | 42 |
| Hα 6562.8 | 1122 | 1155 | 970 | 848 | 600 | 3981(b) | 1245 | 1076 | 754 | 530 | 630 | 1020 |
| [NII] 6585.3 | 206 | 202 | 170 | 122 | 137 | 1498(b) | 445 | 258 | 300 | 276 | 115 | 110 |
| HeI 6678.1 | | 5.8 | 5.7 | 3.9 | | | 5.9 | | | | 3.2 | 10: |
| [SII] 6716.4 | 521 | 130 | 121 | 74 | 40 | 1071(b) | 314 | 253 | 164 | 99.1 | 83.8 | (a) |
| [SII] 6730.8 | 977 | 225 | 220 | 130 | 83 | 2155(b) | 566 | 489 | 318 | 178 | 151 | (a) |
| [SII] 6717/6731 | 0.53 | 0.58 | 0.55 | 0.56 | 0.49 | 0.50 | 0.55 | 0.52 | 0.51 | 0.56 | 0.56 | 0.56 |
| [NII] 5755/6583(c) | 0.109 | 0.11 | 0.13 | | 0.19 | 0.013 | 0.048 | 0.10 | | 0.008 | 0.11 | 0.16 |
| [OIII] 4363/5007 | | | | | 0.083 | | | | | 0.075 | 0.057 | 0.053: |

\* All measurements assume c(Hβ) = 0. See text.
a. Emission line is probably not nuclear: it does not peak at the location of the nucleus and does not match the nuclear Doppler shift. See Fig. 2.
b. These results from Table 2 of TGC93 are discordant with all other measurements of the same lines made by those authors in the east lobe and observations made by others at about the same time. We shall ignore them.
c. This ratio is typically 0.012 in low-density H II regions and PNe. Larger values generally imply densities in excess of the N II $^1D_2$ critical density = $7 \times 10^4$ cm$^{-3}$.